\documentclass[]{aa}  

\usepackage{graphicx}
\usepackage{txfonts}
\usepackage{mathtools}
\begin{document}

   \title{XMM-Newton reveals a Seyfert-like X-ray spectrum in the z=3.6 QSO B1422+231}

%   \subtitle{Have we detected reflection and thermal Comptonization at z=3.6?}

   \author{M. Dadina\inst{1}, C. Vignali\inst{2}, M. Cappi\inst{1}, G. Lanzuisi\inst{2,3}, G. Ponti\inst{4}, B. De Marco\inst{4}, G. Chartas\inst{5}, M. Giustini\inst{6} }

   \institute{INAF/IASF Bologna
              via Gobetti 101, 40129, Bologna, Italy  \email{dadina@iasfbo.inaf.it}
                           \and
 Dipartimento di Fisica e Astronomia dell'Universit\'a degli Studi di Bologna, via Ranzani 1, 40127, Bologna, Italy
 \and
INAF Osservatorio astronomico di Bologna,  via Ranzani 1, 40127, Bologna, Italy
\and
Max-Planck-Institut f{\"u}r Extraterrestrische Physik, Giessenbachstrasse 1, D-85748, Garching, Germany
\and
Department of Physics and Astronomy of the College of Charleston,  Charleston, SC 29424, USA
\and
SRON Netherlands Institute for Space Research, Sorbonnelaan 2, 3584 CA Utrecht, the Netherlands}

   \date{Received  ;  Accepted}
\authorrunning{Dadina et al.}

	\titlerunning{The Seyfert-like X-ray spectrum of z=3.6 QSO B1422+231}

% \abstract{}{}{}{}{} 
% 5 {} token are mandatory
 
  \abstract
   {Matter flows from the central regions of quasars during their active phases are probably responsible for the properties of the super-massive black holes and that of the bulges of host galaxies. To understand how this mechanism works, we need to characterize the geometry and the physical state of the accreting matter at cosmological redshifts, when QSO activity is at its peak.}
   {We aim to use X-ray data to probe the matter inflow at the very center of a quasar at z=3.62. While complex absorption, the iron K emission line, reflection hump and high-energy cut-off are known to be almost ubiquitous in nearby AGN, only a few distant objects are known to exhibit some of them.}
   {The few high quality spectra of distant QSO have been collected by adding sparse pointings of single objects obtained during X-ray monitoring campaigns. This could have introduced spurious spectral features due to source variability and/or microlensing. To avoid such problems, we decided to collect a single epoch and high-quality X-ray spectrum of a distant AGN. We thus picked-up the z=3.62 quasar B1422+231 whose flux, enhanced by gravitationally lensing, is proven to be among the brightest lensed quasars in X-rays (F$_{2-10 keV}\sim$10$^{-12}$erg s$^{-1}$ cm$^{-2}$).}
   {The X-ray spectrum of B1422+231 is found to be very similar to the one of a typical nearby Seyfert galaxy. Neutral absorption is clearly detected (N$_{H}\sim$5$\times$10$^{21}$ cm$^{-2}$ at the redshift of the source) while a strong absorption edge is measured at E$\sim$7.5 keV with an optical depth of $\tau\sim$0.14. We also find hints of the FeK$\alpha$ line in emission at E$\sim$6.4 keV line (EW$\lesssim$70 eV) and a hump is detected in the E$\sim$15-20 keV energy band (rest-frame) in excess of what predicted by a simple absorbed power-law.}
   {The spectrum can be best-modeled with two rather complex models; one assumes the presence of ionized and partially covering matter along the line of sight while the other is characterized by the presence of a reflection component. We argue that the reflection 
seems more plausible here on a statistical basis. In this scenario, the primary emission of B1422+231 is most probably dominated by the thermal Comptonization of UV seed photons in a corona with kT$\sim$40 keV. We also detected a reflection component with relative direct-to-reflect normalization r$\sim$1. 
These findings confirm that gravitational lensing is effective to obtain good quality X-ray spectral information of quasar at high-z, moreover they support the idea that the same general picture characterizing active galactic nuclei in the nearby Universe is valid also at high redshift.}

   \keywords{galaxies: active -- quasar: individual: B1422+231 --X-rays: individual : B1422+231 }
   \maketitle
%

%________________________________________________________________

\section{Introduction}

%The presence of matter close to the central super-massive black-holes (SMBH) 
%hosted by active galactic nuclei (AGN) appears in the X-ray 

High-energy observations are used to study the geometry and physical condition of matter in the inner regions of active galactic nuclei (AGN), since X-rays are thought to originate from close to the central supermassive black hole ((SMBH, see Antonucci 1993; Fabian 2000). In nearby Seyfert or radio galaxies, in fact, absorption due to cold matter along the line of sight imprints a low energy cut-off in their X-ray spectra (Smith \& Done 1996; Bassani et al. 1999, Cappi et al. 2006, Dadina 2007, 2008). In addition, most AGN observed with sufficient statistics also reveal the presence of a smoothed spectral curvature due to warm  matter below $\sim$4 keV (Piconcelli et al. 2005). X-ray spectra of local AGN often display emission features which are diagnostic of both the geometry of the matter infalling onto the SMBH and the emission mechanism acting in producing the high-energy emission itself (Nandra \& Pounds, 1994; Smith \& Done, 1996; Perola et al. 2002; Cappi et al. 2006; Dadina 2008; Malizia et al. 2014; Fabian et al. 2015). The most prominent of these features is the iron line at 6.4-6.9 keV. It may appear shifted in energy and skewed by relativistic effects whose strength depends on the vicinity of the line production region to the SMBH (Tanaka et al. 1995; see, Fabian et al. 2000 for a review on the topic). In addition, a hump  due to the reflection of the primary emission by the matter surrounding the SMBH (Lightman \& White 1988; Guilbert \& Rees 1988) is usually seen at E$\sim$20-40 keV (Perola et al. 2002;  Dadina 2008). Broad band  (0.1-200 keV) X-ray observations of nearby AGN allowed also to measure the presence of a high-energy cutoff E$_C$ at E$\sim$70-200 keV (Perola et al. 2002;  Dadina 2008; Malizia et al. 2014; Fabian et al. 2015). This feature is expected in the so called ``two-phase models'' for the production of the high-energy photons in radio-quiet AGN. In this scenario, optical-UV seed photons coming from the accretion disk are Comptonized into the X-ray band by the electrons forming a hot corona that sandwiches the accretion disk itself. The electrons have a thermal energy distribution and the shape of the emerging X-ray spectrum displays a high-energy cut-off at E$_{cut-off}\sim$2$\times$kT$_{e^{-}}$ (Haardt \& Maraschi 1993; Haardt, Maraschi \& Ghisellini 1994; Petrucci et al. 2001).

At intermediate (z$\geq$0.1) and high (z$\geq$1) redshift, X-ray spectra of QSO often display only absorption and/or FeK$\alpha$ line in emission (Vignali et al. 2006), the latter being sometimes associated with Compton-thick absorbers (Vignali et al. 2010, 2014). In the framework of unified models (UM) for AGN (Antonucci 1993), at least at the zero-th order, the only differences between low- and high-z AGN should be due to cosmological evolution of the environment in which the SMBHs are embedded. Some confirmation of this picture have been effectively obtained. No strong evidence of evolution of the intrinsic X-ray properties of QSO with redshift has been observed (Vignali et al. 2005, but see Saez et al. 2008 for a suggested limited evolution of the photon index with z while the measurement of a different evolution between absorbed and non-absorbed sources (Aird et al. 2015) seems to indicate that the circumnuclear environment may evolve, as expected, with z. Similarly, the presence of a strong correlation between the X-ray photon index and the Eddington luminosity of quasars (Lu \& Yu 1999, Shemmer et al. 2006, 2008; Risaliti, Young \& Elvis 2009; Fanali et al. 2013; Brightman et al. 2013) has been explained in the framework of the two-phase model for the production of X-rays (Pounds, Done \& Osborne 1995) thus indicating that it acts both in the nearby and distant Universe. Furthermore, there have been recent claims that reflected and relativistically blurred components are also present in the X-ray spectra of some high-z (z$\geq$2) and lensed QSO after stacking many and sparse X-ray observations of each sources (Reis et al. 2014; Reynolds et al. 2014; Walton et al. 2015, but see also Chartas et al. 2007, Chartas et al. 2016). Moreover, transient and variable absorption iron lines linked to hot and fast outflowing gas (ultra fast outflows, aka UFO) and with quite similar characteristics were observed in the rest frame 7-10 keV spectra of both nearby Seyferts (Pounds et al. 2003; Dadina et al. 2005; Tombesi et al 2010a, 2010b; Gofford et al. 2013) and in an handful of bright/distant (z$\geq$1) QSO (Chartas et al. 2002, 2003, 2007, 2014; Lanzuisi et al. 2012; Vignali et al. 2015). 

The above mentioned evidences are consistent with the idea that the emission mechanism does not dramatically change over cosmic time, while some evolution with redshift may be present for the properties of the matter flows around SMBH. High signal-to-noise X-ray spectra are fundamental to fully characterize the environment in the central engines of AGN at high-redshift and how they evolve throughout  cosmological time. Since we measure a tight correlation between the mass of the galaxies bulges and the mass of their central SMBH (Kormendy \& Richstone, 1995; Magorrian et al. 1998; Ferrarese \& Merritt, 2000; Merritt \& Ferrarese, 2001; Marconi et al. 2004), we now know that a feedback mechanism must be at work. However, to fully understand how this works, we must constrain the physical conditions of the objects in epochs where both QSO and starburst activity were at their highest (Boyle et al. 1988; Madau et al. 1996; Brandt \& Hasinger 2005 and references therein). For example, the measurement and characterization of UFO at high- z is of crucial importance to understand when and how the feed- back mechanism between the SMBH and the host galaxy bulges has set-up and how it has evolved along cosmic time.

   \begin{figure}
   \centering
   \includegraphics[width=6.0cm, height=8.0cm, angle=-90]{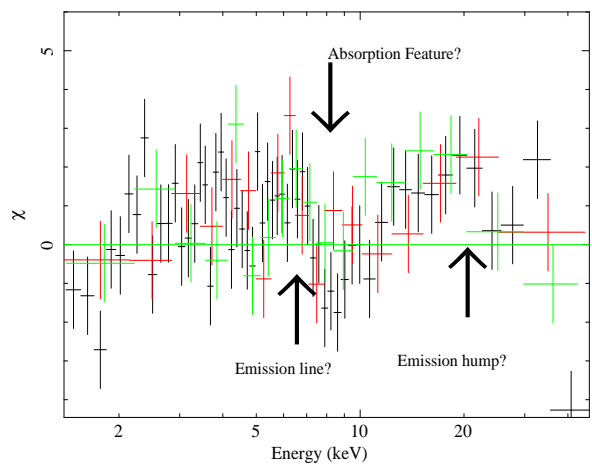}
      \caption{Data-to-model ratio expressed in terms of standard deviations once the X-ray data are fit with a simple power-law model absorbed by the Galactic column of N$_H$=3.2$\times$10$^{20}$ cm$^{-2}$ (Kalberla et al. 2005). Data have been rebinned so as to have 20$\sigma$ significance points for clarity purposes. EPIC-pn data points are in black, while MOS1 and MOS2 data points are in red and green respectively.}
   \end{figure}

 The spectroscopic X-ray studies that allowed the detection of UFO and relativistically blurred features of the most distant objects mentioned above, took advantage of the fact that all the studied QSO, except for  HS 1700+6416 and PID352 (Lanzuisi et al. 2012; Vignali et al 2015), are lensed. Thus they display magnified emission and this allows the collection of X-ray spectra of higher signal-to-noise (S/N). While waiting for the next generation of large collecting area X-ray telescopes such as Athena (Nandra et al. 2013), the usage of ``cosmic lenses'' may play a crucial role to extend at high redshift studies that are up to now limited to the nearby Universe.  We present here the first high-quality X-ray spectrum, obtained with a single XMM-Newton observation of B1422+231, of a lensed and radio-loud (R$_{L}\sim$300, but see section 3.1 for a discussion on this point) QSO at z=3.62 (Patnaik et al. 1992; Misawa et al. 2007). Its four images are all within $\sim$5'', thus the source is ``point-like'' for XMM-Newton, and the magnification factor has been estimated to be between $\mu\sim$15-76 (Kormann et al. 1994; Chiba 2002; Raychaudhury et al. 2003; Assef et al. 2011). For this work we assume a magnification factor of  $\mu\sim$15-20, a range of values that is common for the majority of the works presented above. A broad H$\beta$ emission line has been observed in its optical spectrum and the mass of the central SMBH has been estimated to be $\sim 5\times10^{9} M\sun$ using single epoch CIV measurements (Greene et al. 2010) and later confirmed using also other optical lines (Assef et al. 2011). The X-ray spectrum of the source was previously presented in Misawa et al (2008) that analyzed poor quality spectra obtained with Chandra and XMM-Newton\footnote{Chandra pointed the source three times collecting 426, 244 and 762 counts respectively. XMM-Newton observed the source for $\sim$5 ks only and the source fell on a EPIC-pn chip gap precluding the use the pn data. The two EPIC-MOS collected $\sim$195 counts each (Misawa et al. 2008).}. In these low S/N spectra, B1422+231 displayed rather typical and almost featureless X-ray spectra with a photon index $\Gamma\sim$1.6, with only hints of absorption (N$_{H}\leq$10$^{22}$ cm$^{-2}$) at low energies and an observed and integrated flux of F$_{2-10 keV}\sim$9$\times$10$^{-13}$ erg s$^{-1}$ cm$^{-2}$.

\section{Data reduction and analysis}

XMM-Newton pointed at B1422+231 on July 31, 2014. The data reduction has been performed using the package SAS-14. The high background periods have been cleaned by excluding the first 2 ks of the pointing, so as to eliminate the first high peak of soft protons and then applying a filter on the count-rate (0.65 and 0.2 c/s for pn e MOS respectively). After this filtering, we obtained a net exposure of $\sim$65, 74.2 and 64.3 ks with pn, MOS1 and MOS2 respectively. Source counts have been extracted from circular regions with radii of 30 and 40 arcsec for pn and MOS detectors respectively allowing to collect 25.8, 8.2 and 7.8 k-counts with pn, MOS1 and MOS2 respectively. Background counts were extracted from source free circular regions larger than that centered on the source but in the same chip of the source.

No variability has been detected at more than 75-80\% significance level and the observed average flux is F$_{2-10 keV} \sim 10^{-12}$ erg s$^{-1}$ cm$^{-2}$ once the data are fitted with a power-law absorbed by the Galactic column density (model \#1 in Table 1). This yields an observed integrated luminosity L$_{2-10 keV} \sim$6$\times10^{46}$ erg s$^{-1}$ (we use H$_0$=70 km s$^{-1}$ Mpc$^{-1}$, $\Omega_M$=0.286 and $\Omega_{\lambda}$=0.714). Errors are reported at 90\% confidence level in table \#1 and throughout the paper.

 \begin{table}
\scriptsize
      \caption[]{Phenomenological models: Column 1: Model number; Column 2: absorbing column in excess of the Galactic value; Column 3: photon index; Column 4: energy centroid of the emission Gaussian line (E$_{edge}$ in the lower row of model \#3); Column 5: Line width (1$\sigma$) ($\tau_{edge}$ in the lower row of model \#3 ; Column 6: Line equivalent width; Column 7: $\chi^2$/degrees of  freedom.}
         \label{tab1}
         \centering
         \begin{tabular}{lcccccc}
 \hline\hline
  &&&&&&\\
  & N$_{H}$         & $\Gamma$ & E$_{FeK\alpha}$ & $\sigma_{FeK\alpha}$ & EW$_{FeK\alpha}$& $\chi^2$/d.o.f.\\
  &&&(E$_{edge}$)&($\tau_{edge}$)& &\\
      & 10$^{21} cm^{-2}$ &          & keV (keV)           &  eV            &  eV          &               \\
  &&&&&&\\
\hline
  &&&&&&\\
1  &                   &1.61$^{+0.02}_{-0.02}$& & & &1282/1189\\
  &&&&&&\\
2  & 3.42$^{+1.12}_{-1.11}$&1.66$^{+0.02}_{-0.02}$&6.35$^{+0.18}_{-0.22}$&$\leq$0.48&62$^{+48}_{-34}$&1256/1185\\
  &&&&&&\\
3 &2.37$^{+0.10}_{-0.10}$&1.65$^{+0.02}_{-0.02}$&6.38$^{+0.13}_{-0.20}$&$\leq$0.38&$\leq$69&1228/1183\\
  &&&&&&\\
  &&&(7.48$^{+0.20}_{-0.23}$)&(0.14$^{+0.04}_{-0.04}$)&&\\
  &&&&&&\\
  \hline
\end{tabular}

\normalsize

\end{table}

The spectral analysis of the source has been performed averaging the data over the entire good time exposure and grouping the data so as to have 20 counts in each spectral bin. As a sanity check, however, all the results presented below have been tested also using data collected only in the first, middle and last third of the observation and no differences were found. This has been done to detect, if any, evidences of time-dependent spectral variations that, in principle, could be related to either the intrinsic source's variability or the ``microlensing'' phenomenon (Paczynski 1986; Wambsganss 1990).

   \begin{figure*}
   \centering
   \includegraphics[width=6.3cm, angle=-90]{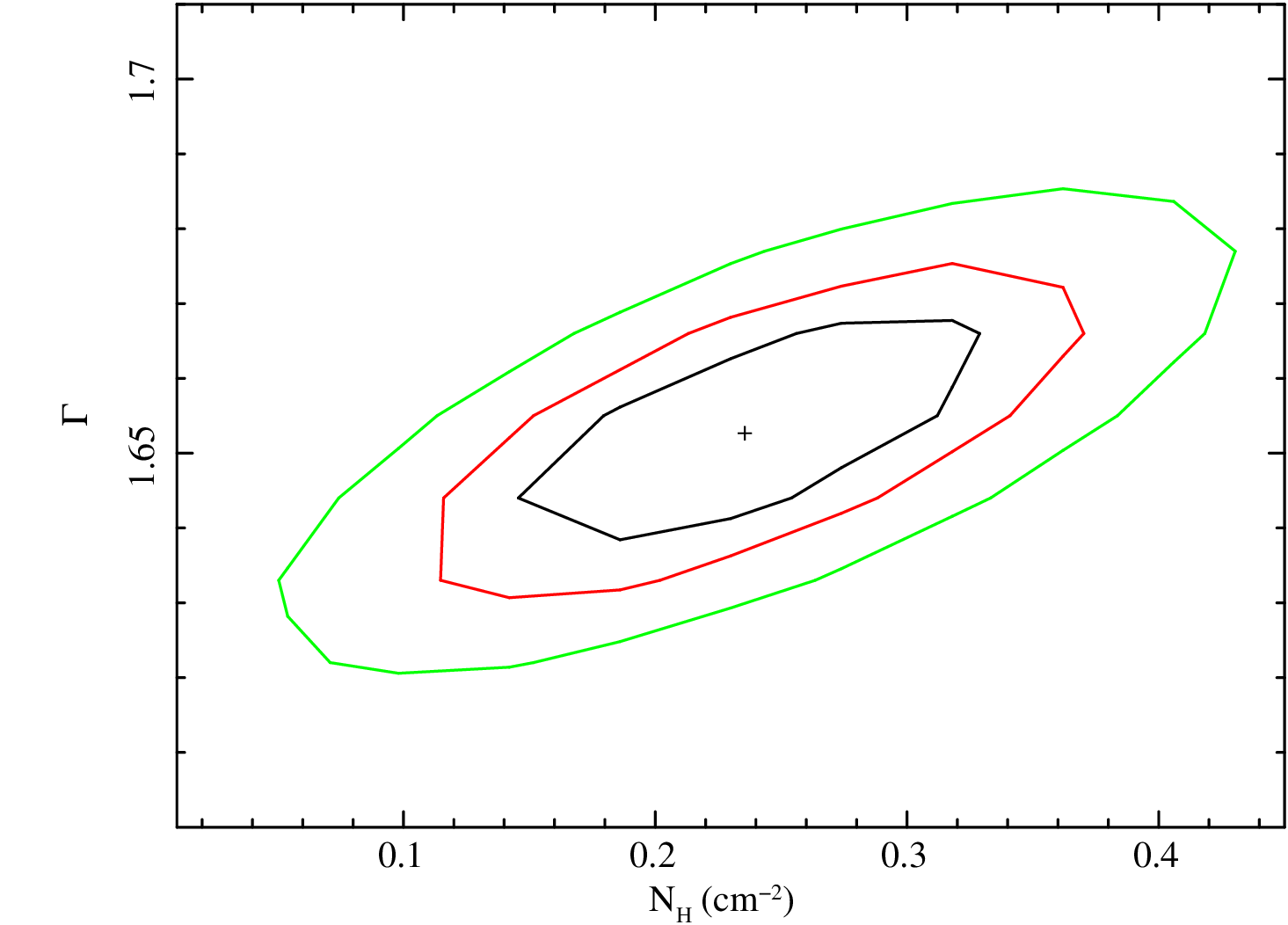}\includegraphics[width=6.3cm, angle=-90]{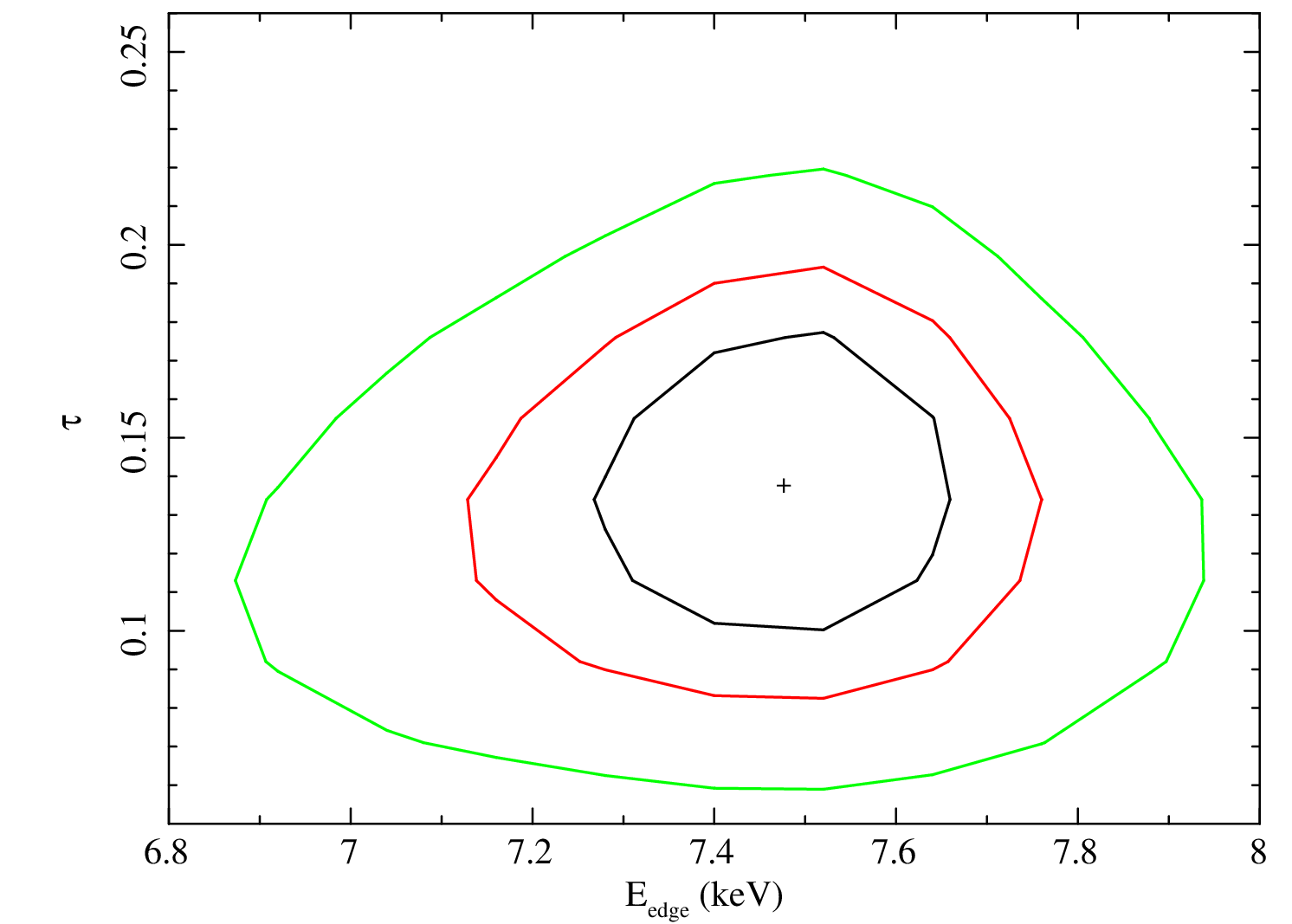}

\includegraphics[width=6.3cm, angle=-90]{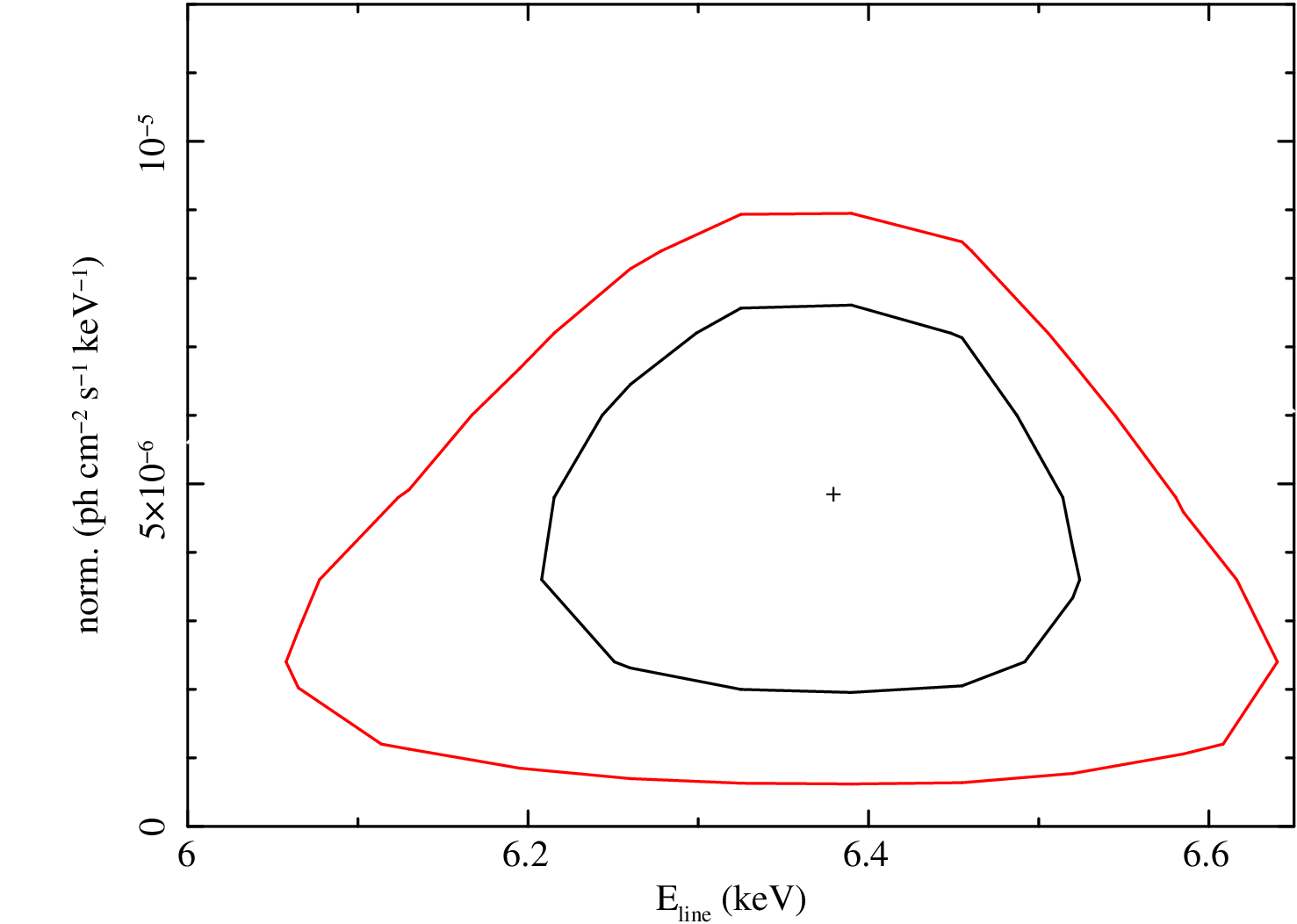}\includegraphics[width=6.3cm, angle=-90]{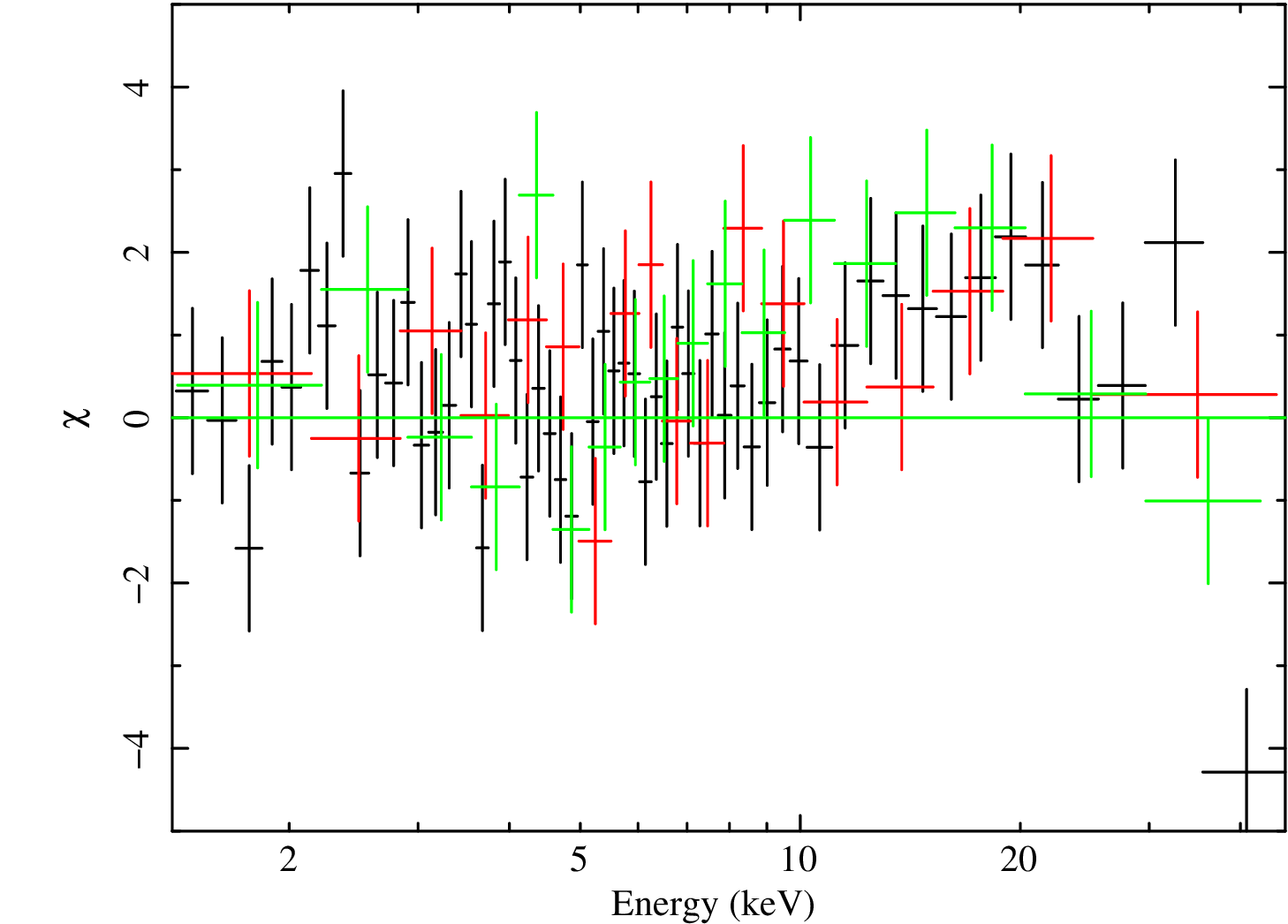}
      \caption{99, 90 and 68\% confidence contours plots obtained using a model composed of a primary power-law absorbed by a cold column and by an edge at E$\sim$7.5 keV plus a Gaussian in emission (model \#3 in Table 1). From left to right: ({\it upper-left panel}) photo-index $\Gamma$ vs absorbing column density (N$_{H}$ in units of 10$^{22}$); ({\it upper-right panel}) optical depth of the edge ($\tau$) vs. its energy centroid; ({\it lower-left panel}) Gaussian emission line normalization vs. its Energy. In this case only 90 and 68\% confidence contours are presented since the emission Gaussian line is detected with $\sim$97\% confidence in this model and the 99\% confidence contour gives only an upper limit; ({\it lower-right panel}) data-to-model ratio expressed in term of standard deviations. Binning and colors as in figure 1.}
         \label{FigVibStab}
   \end{figure*}

In Fig. 1 the data-to-model ratio expressed in terms of standard deviations is shown (rest-frame energy). The data are fitted using the model \#1 of table 1 which consists of a simple power law absorbed (phabs model in Xspec) by Galactic column density (N$_H$=3.2$\times$10$^{20}$ cm$^{-2}$, Kalberla et al. 2005). The X-ray spectrum of the source clearly displays three main features: 1) a low energy cut-off due to an absorption component in excess of the Galactic one; 2) a dip in counts at E$\sim$7-8 keV plus the possible presence of a feature in emission at  E$\sim$6.5 keV; 3) an excess-hump at E$\sim$15-30 keV.

\begin{table*}

\small

\begin{flushleft}
\vspace{0.2cm}
\caption[]{Best fit models}

{\bf Complex absorber model:} Column 1: Model name; Column 2: absorbing column in excess to the Galactic value; Column 3: photon index; Column 4: column density of the warm absorber; Column 5: covering factor of the warm absorber; Column 6: ionization parameter of the warm absorber; Column 8: energy of the Gaussian line in emission (rest frame); Column 8: width of the Gaussian line in emission (rest frame); Column 9: equivalent width of the Gaussian line in emission (rest frame);  Column 10: $\chi^2$/degrees of  freedom.
\end{flushleft}

\vspace{0.1cm}

         \centering
         \begin{tabular}{lccccccccc}
 \hline\hline
  &&&&&&\\
Complex absorber   & N$_{H,cold}$    & $\Gamma$  & N$_{H, warm}$ &                C$_f$ & Log($\xi^{a}$) & E$_{FeK\alpha}$ & $\sigma_{FeK\alpha}$ & EW$_{FeK\alpha}$ & $\chi^2$/d.o.f.\\
  &&&&&&\\
      & 10$^{21} cm^{-2}$ &      &   10$^{21} cm^{-2}$        &    &   &  keV      &  keV    &   eV     &            \\
  &&&&&&\\
\hline
  &&&&&&&&&\\
  & 8.84$^{+2.16}_{-2.15}$&1.95$^{+0.09}_{-0.09}$&961$^{+11}_{-12}$&0.43$^{+0.08}_{-0.09}$&2.00$^{+0.21}_{-0.58}$&6.36$^{+0.36}_{-0.32}$&$\leq$822  & 38$^{+30}_{-30}$ & 1207/1182\\
  &&&&&&&&&\\
  \hline
\end{tabular}

\begin{flushleft}
 
$^{a}$ $\xi$ expressed in erg s$^{-1}$ cm  

\vspace{0.4cm}
     {{\bf Reflection model:} Column 1: Model name; Column 2: absorbing column in excess to the Galactic value; Column 3: photon index; Column 4: High-energy cut-off (rest frame); Column 5: relative reflected-to-direct emission normalization; Column 6: inclination angle; Column 7: $\chi^2$/degrees of  freedom.}
\end{flushleft}

         \label{tab1}
         \centering
         \begin{tabular}{lcccccc}
 \hline\hline
  &&&&&&\\
Reflection dominated   & N$_{H}$    & $\Gamma$  & E$_{cut-off}$& r &$\theta$& $\chi^2$/d.o.f.\\
  &&&&&&\\
      & 10$^{21} cm^{-2}$ &      &   keV&    &    degrees &            \\
  &&&&&&\\
\hline
  &&&&&&\\
    & 4.57$^{+1.79}_{-1.73}$&1.81$^{+0.08}_{-0.07}$&80$^{+34}_{-19}$&1.10$^{+0.31}_{-0.31}$&45$^{fixed}$& 1207/1186\\
  &&&&&&\\
  \hline

\end{tabular}

\normalsize

\end{table*}

To perform phenomenological tests on the presence of these features  we first added an in-situ (z=3.62) extra-column in absorption and a Gaussian line  in emission (model \#2 in Table 1), since they are known to be common in the X-ray spectra of AGN (Nandra \& Pounds 1994; Smith \& Done 1996; Vignali et al. 2006). We obtained a $\Delta\chi^2\sim$15 for a single parameter of interest when a cold absorber is added to model \#1 of Table 1 and  $\Delta\chi^2\sim$9 for three parameters of interest for the Gaussian line in emission. Moreover, to test the significance of the count drops at E$\sim$7.5-8 keV, we used an absorption edge (model \#3 in Table 1). Overall we obtain that the simultaneous detection of an intrinsic absorber and of an edge is significant at more than 99\%\footnote{The intervals of confidence reported here are calculated using the F-test which is known to be inaccurate and, in particular, slightly optimistic (Protassov et al. 2002). Here we use the F-test probabilities as mere indications of the strength of the putative spectral features.} confidence level ($\Delta\chi^2\sim$28 with respect to model \#2 of Table 1, see also the upper panels of Fig.2)  while the strength of the detection of a Gaussian line in emission depends on the presence of the absorption edge in the underlying model: the line in emission  is detected at high significance when the dip of counts at $\sim$7.5 keV is not modeled (model \#2 in Table 1) but it is detected at only $\sim$97\% confidence level when the absorption edge is modeled (see lower-left panel in Fig. 2 and the parameters of model \#3 in Table 1). In any case, if the width of the line is left free to vary, we obtain that it is consistent with being narrow. 

The parameters obtained with this simple and phenomenological modeling of the data are well in agreement with what previously presented by Misawa et al. (2008) who analyzed X-ray spectra with few hundreds of counts. In particular, the high-energy continuum is recorded to be rather flat ($\Gamma\sim$1.6) and absorbed by a column density of cold matter with N$_H$$\sim$2.4$\times$10$^{21}$ cm$^{-2}$. Both these values are well within the range and upper limits previously reported (Misawa et al. 2008).  It is worth considering, however,  that all these features are unable to account for the hump at energies higher than 10 keV (lower-right panel of Fig. 2) and this indicates that more complex modeling of the data is needed.

Finally, to further test the possible presence of the above mentioned spectral features, we analyzed the RGS data using standard techinques. The main goal was to test the intrinsic width of the emission line. However, we obtained a poor quality RGS spectrum (we obtained $\sim$5000 source plus background counts, only $\sim$20\% of which due to the source) that is plotted in Fig. 3. Using model \#1 in table 1 as undelying continuum, we thus checked that the intensity of emissione line and the optical depth of the absorption edge are consistent with what obtained with EPIC instruments. No further constraints on the parameters of the emission line are obtained.

\subsection{A complex absorbing system?}

To simultaneously model the detected features we first assumed an absorption dominated scenario. The absorption edge measured at E$\sim$7.5 keV, in fact, may indicate that the primary emission of B1422+231 intercepts along the line of sight some cold/ionized matter that only partially covers the central engine of the source. This would introduce an underestimate of the total absorber and, thus, the appearance in the data of the spurius edge at E $\geq$7.1 keV. To study this scenario, we used the {\it zxipcf} model (Reeves et al. 2008) within $Xspec$ which is able to well reproduce the data (see left panel of Fig. 4 and the upper panel of Table 2) once it is combined to cold absorber (see table 2). We also used pre-compiled Xstar tables, namely mtab1e 8, 21, 22 and 25 (each obtained using a ionizing flux with power-law spectra with $\Gamma$=2, i.e. similar to what obtained here in the ionized absorption scenario)  that have turbulence velocities from 100 to 3000 km/s and cover large ranges of absorbing columns (N$_H\sim$10$^{20-24}$ cm$^{-2}$). We tested both partial and complete absorption scenarios never obtaining spectral fits with $\chi^2$ values better than those obtained with {\it zxipcf}.

   \begin{figure}
   \centering
   \includegraphics[width=6.0cm, height=8.0cm, angle=-90]{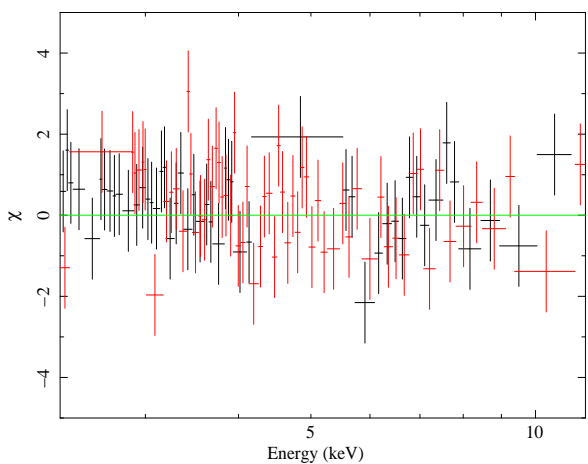}
      \caption{Rest frame data-to-model ratio expressed in terms of standard deviations  once the RGS data are fit with model \#1 of table 1. Data have been rebinned so as to have 20 source plus cunts per energy bin. RGS1 and RGS2 datapoints are in black and red respectively.}
   \end{figure}

The FeK$\alpha$ (E$_{FeK\alpha}\sim$6.4 keV) line is only marginally detected in this case with a 90\% upper limit of EW$\lesssim$70 eV (rest-frame). This value is well in agreement with what is expected if the line is produced in the cold absorber itself (EW$\sim$30 eV, Makishima 1986; Leahy \& Creighton, 1993). It is also consistent with what is measured for the broad component of the FeK$\alpha$ line in nearby type I AGN (EW$\sim$75 eV, de La Calle P{\'e}rez et al. 2010) or nearby radio-galaxies (Grandi \& Palumbo 2004, 2007).  
Nonetheless, it is slightly lower than what is expected if the presence of a standard accretion disk in the inner core of B1422+231 is assumed (EW$\sim$100-120 eV for type I AGN, Matt, Perola \& Piro, 1991).  
The edge and the hump at energies above 10 keV are here accounted for by the presence of the partially covering and warm component with high column density (N$_H\sim$10$^{24}$ cm$^{-2}$, Log($\xi$  erg s$^{-1}$ cm)$\sim$2 and C$_f\sim$0.4). In the left panel of Fig. 4 the data-to-model ratio expressed in terms of standard deviation as a function of rest-frame energy are presented. Some features may still be present at E$\sim$15 keV (rest frame). 
It is worth noting, however that in the observed frame this corresponds to E$\sim$2-3 keV, i.e. an energy band where instrumental absorption edges could well be important\footnote{see XMM-Newton Users' Handbook at http://xmm.esac.esa.int } (but see the end of sect. 2.2 for a more detailed discussion on this topic).

   \begin{figure*}
   \centering
   \includegraphics[width=6.5cm, angle=-90]{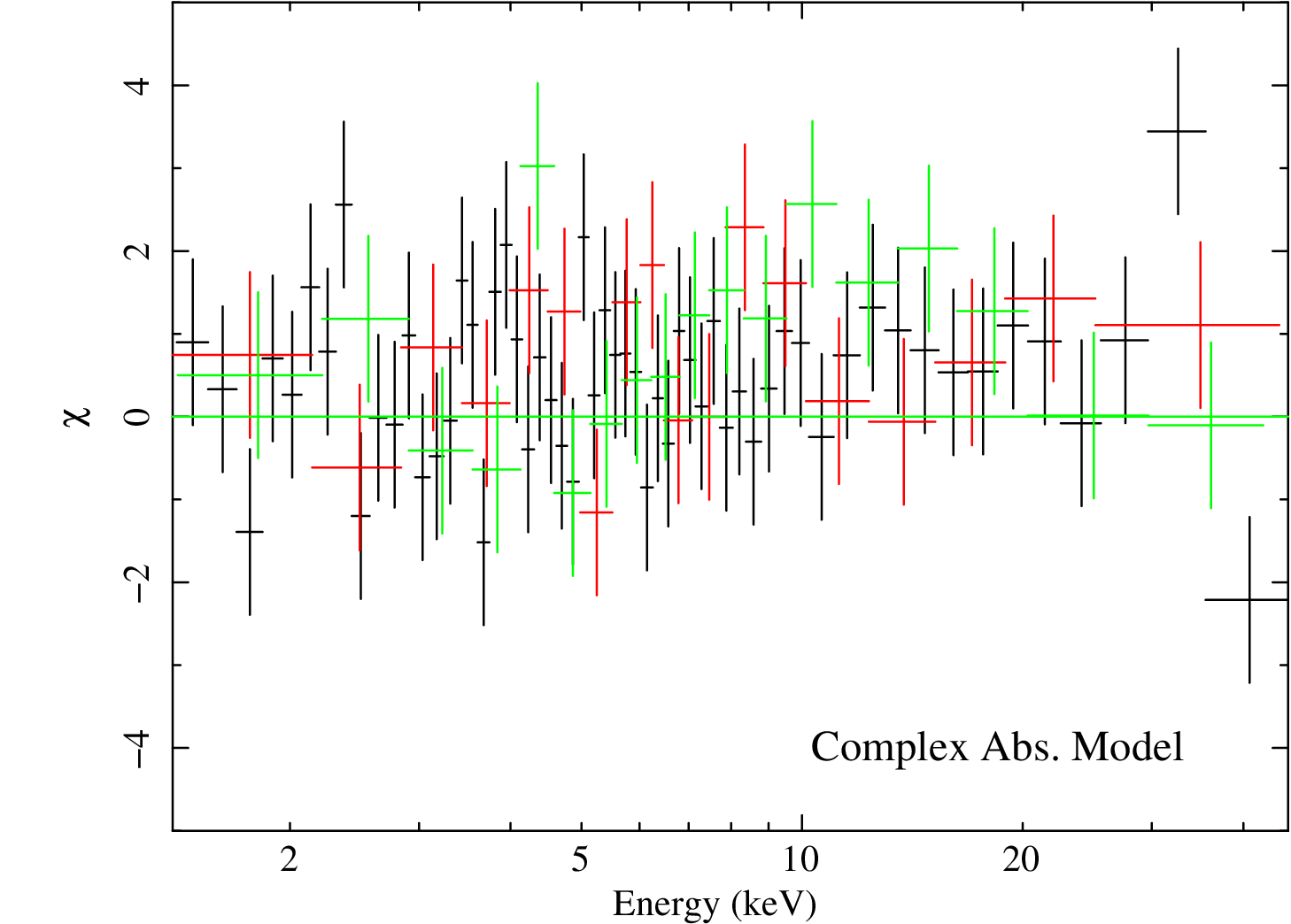}
   \includegraphics[width=6.5cm, angle=-90]{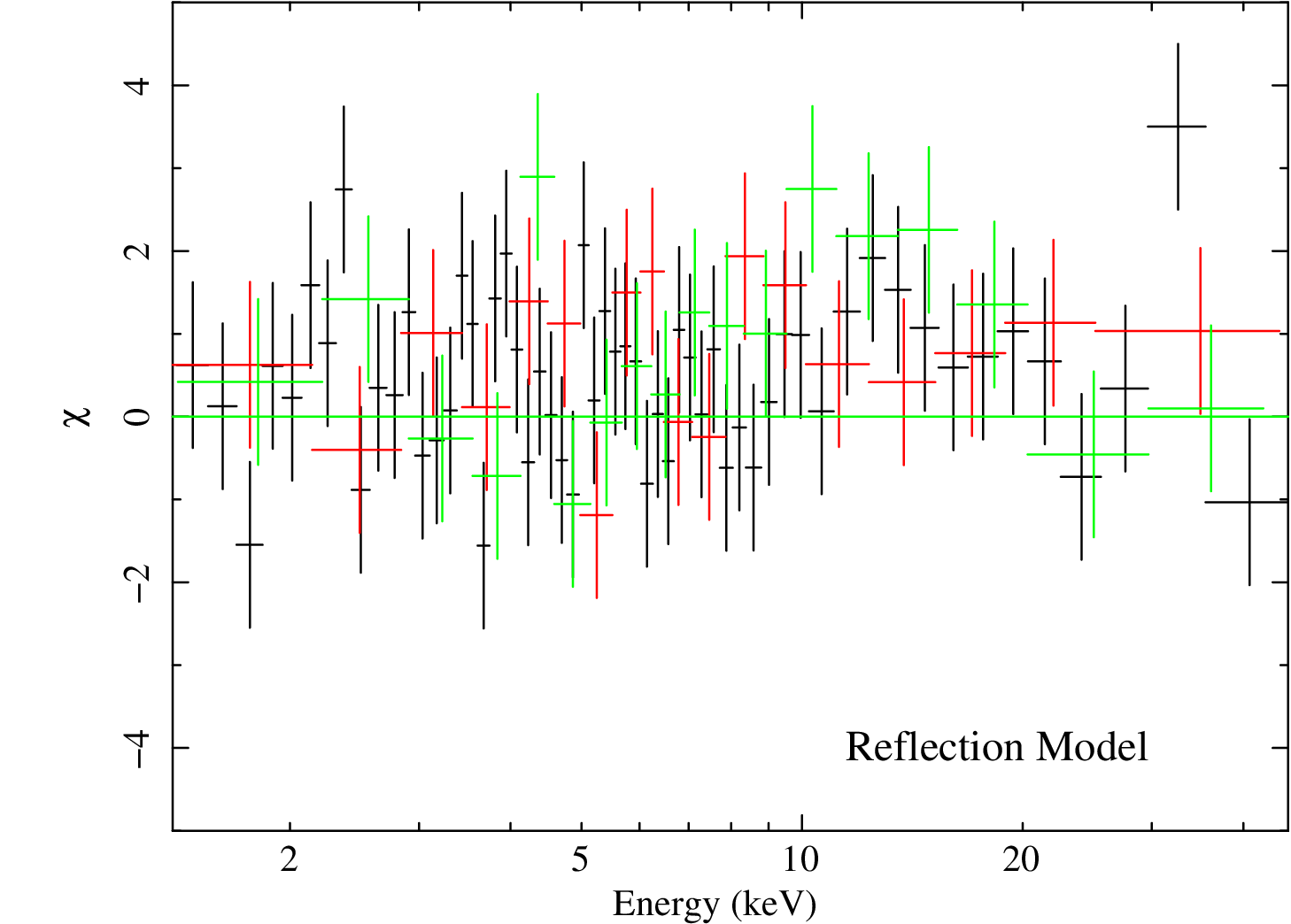}
      \caption{Data-to-model ratios in terms of standard deviations when the data are fitted by  a complex absorption a reflection (left panel) and a reflection model (right panel). Binning and colors as in Figure 1.}
         \label{FigVibStab}
   \end{figure*}

\subsection{Is a reflection component present?}

Both the $\sim$7.5 keV edge and the $\sim$15-20 keV hump could be due to the presence in the X-ray spectrum of B1422+231 of a reflection component. To test its presence we used the $Pexmon$ model in $Xspec$ (Nandra et al. 2007) which consistently models the continuum hump at energies above 10 keV and the associated emission lines\footnote{See also Murphy \& Yaqoob 2009 for a discussion on the possible underestimation of the reflection component once assuming slab geometries, as done in $pexmon$, for the reflector}. Given the limited statistics available, we fixed some of the variables of the model: the inclination angle of the reflecting system with respect to the line of sight was fixed to $\Theta$=45$^{\circ}$ and the abundance of all elements were set equal to solar (once left free to vary, the inclination angle was statistically unconstrained, while the abundance was found to be larger than $\sim$0.3 solar at 90\% confidence level with a $\Delta\chi^2\sim$0.5 for one more parameter of interest and for a best value Abund$\sim$0.95 solar). 
As reported in the lower panel of table 2, this model provides an excellent fit of the data (see also right panel of Fig. 4). As often observed in nearby Seyfert galaxies and radio galaxies, after the addition of the reflection component, the slope of the primary power-law steepens reaching values ($\Gamma\sim$1.8) commonly observed in the nearby Universe (Dadina 2008, Cappi et al. 2006) and perfectly consistent with what is predicted in the two-phase scenario (Haardt, Maraschi \& Ghisellini, 1997). Moreover, taking advantage of the large redshift of B1422-231, we are able to measure a high-energy cut-off (E$_c\sim$80 keV, see Table 2 and left panel of Fig. 5). The reflection component is measured at a good significance level (right panel of Fig. 5), and the relative intensity of the reflected component with respect to the direct continuum is r$\sim$1, i.e. consistent with a 2$\pi$ coverage of the reflector. We also tested the hypothesis of relativistically blurred reflection by smearing the $Pexmon$ component with a relativistic kernel (namely kdblur in Xspec) but the data did not  statistically require the addition of such effect.

   \begin{figure*}
   \centering
   \includegraphics[width=6.5cm, angle=-90]{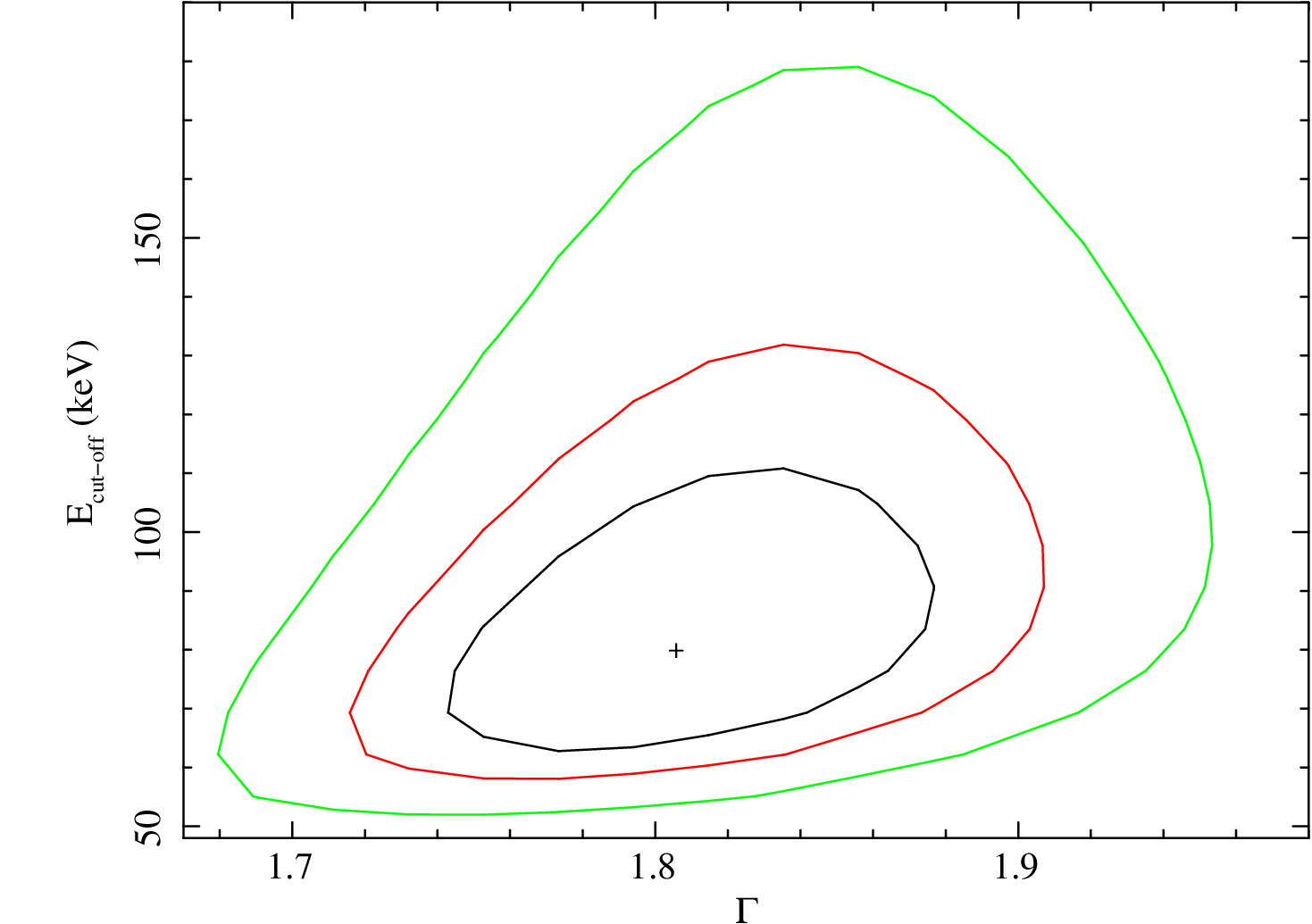}\includegraphics[width=6.5cm, angle=-90]{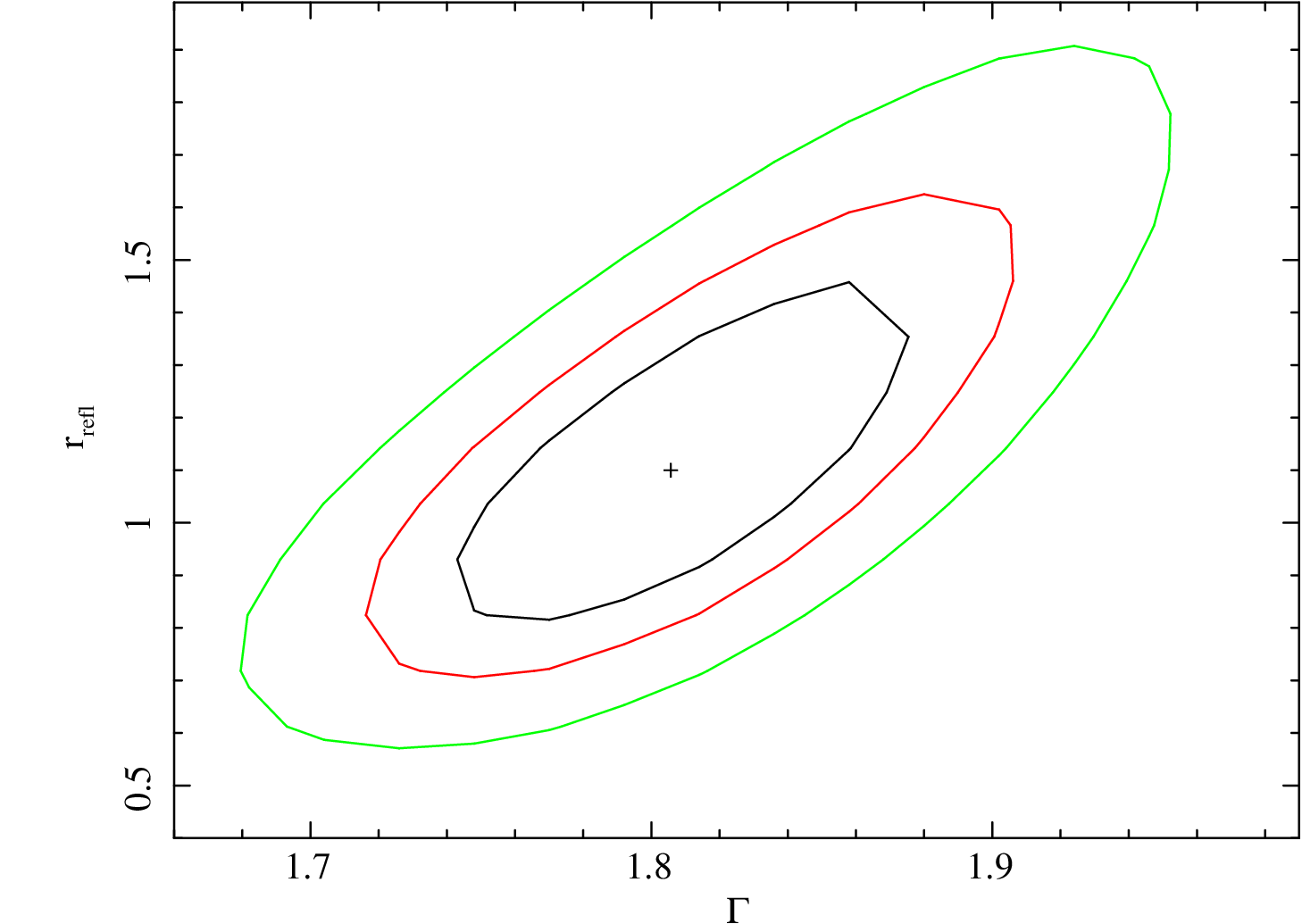}
      \caption{Confidence contours of the high-energy cut-off vs. photon index (left panel) and relative normalization of the reflected component vs photon index (right panel) as measured when the X-ray data of B1422+231 are modeled using $Pexmon$ (Nandra et al. 2007).}
         \label{FigVibStab}
   \end{figure*}

As for the case of complex absorber, some residuals remain at $\sim$15 keV  even in the reflection scenario (see right panel of Fig. 4). We first tried to test its statistical importance. Since the residuals in the reflection scenario seem more peaked than in the complex absorption case, we first added a broad Gaussian in emission and found that the fit improvement is poor  ($\Delta\chi^2\sim4$ for three more parameters of interest). This indicates that the addition of such feature is not, statistically speaking, required by the data. The energy of the line, moreover, is found to be E$\sim$2.7 keV (observed frame) and this suggests that at least part of the residuals may be due to problems of calibrations around the AU M edge due to the mirror coating (E$\sim$2.5 keV). On more robust physical basis, we also tested the addition of another reflection component due, for example, to the contribution of the putative torus and/or accretion disk, without finding any significant improvement in the fit. If we check for the ionization state of reflector using XILLVER model in Xspec (Garc\'{i}a et al. 2014) we find that the ionization parameter is consistent with zero and lower than Log($\xi$  erg s$^{-1}$ cm)$\leq$2.2. We finally stress here that some sort of ``spectral noise'' may be also introduced by the fact that we are observing a quadruple lensed object. This means that the light paths of the four images may be different and for this reason each of the four images may show slightly different X-ray spectra.

\subsection{Searching for hints of outflowing matter}

   \begin{figure}
   \centering
   \includegraphics[width=5.5cm, angle=-90]{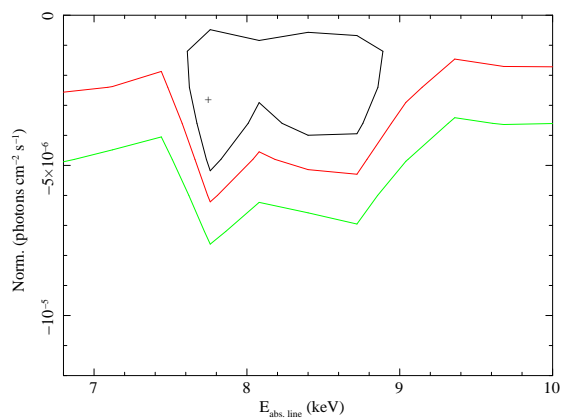}
      \caption{Confidence contours of the normalization of a Gaussian line in absorption vs its energy centroid. The baseline model is here the reflection dominated one presented in the lower panel of table 2. The inclusion of the absorption line yields to $\Delta\chi^2\sim$2.5 for two more parameters of interest. The line width has been fixed to $\sigma$=10 eV.}
         \label{FigVibStab}
   \end{figure}

We also tested, both in the absorption and reflection scenarios the possible presence of out-flowing matter that could be responsible for blue-shifted absorption at energies close to that of the absorption edge. Following Tombesi et al. (2010) and Chartas et al. (2003) we first modeled this component with a simple Gaussian in absorption. The method consists in performing a blind search over the entire X-ray spectrum of a narrow absorption Gaussian line and then checking their statistical strength by performing Monte-Carlo simulations. The simulations always found that the significance of the absorption lines is slightly lower than that calculated with the F-test (Tombesi et al. 2010). Here we adopt a simplified version of this procedure, without Monte-Carlo simulations, since we found that the feature is, statistically speaking, not required by the data (see Fig. 6). We had a similar result adopting a P-Cygni profile to simultaneously model the emission line and the absorption edge as recently observed in PDS 456 (Nardini et al. 2015). In this case we used the model by Done et al. (2007). This component, when added to the simply absorbed power-law is not able to model all the spectral complexity observed between 6-8 keV (rest-frame) while, when added in the complex absorption and reflection scenario, it does not allow to significantly improve the modeling of the data ($\Delta\chi^2\lesssim2.5$ for three more parameters of interest when the P-Cygni profile model is added to the complex absorber scenario).

\section{Discussion}

In this paper we presented the complex X-ray spectrum of the distant (z=3.62) and lensed QSO B1422+231. The magnification factor $\mu\sim$15-20 (Kormann et al. 1994, Assef et al. 2011) and the long XMM-Newton pointing allowed us to obtain a very high quality X-ray spectrum considering the distance of the target (D$_L$=32.2 Gpc). The XMM-Newton pointing, in fact, allowed to collect in a single observation $\sim$40 kcounts (considering the three EPIC instruments on board XMM-Newton) in the 0.3-10 keV (observer frame) energy range. This is one of the best single epoch X-ray spectrum of high-z QSO ever obtained. To our knowledge, the only observations of high redshift QSO of comparable quality are those obtained with XMM-Newton for the lensed QSO APM 08279+5255 (z=3.92, see Saez \& Chartas 2011 for a list of X-ray observations of the source) with $\sim$16.5 kcounts for their deepest XMM-Newton observation. 

The integrated X-ray flux recorded during the $XMM$-$Newton$ pointing presented here is F$_{2-10 keV}\sim$$10^{-12}$ erg s$^{-1}$ cm$^{-2}$ that corresponds to an unobscured luminosity of L$_{2-10 keV}\sim$7.3$\times$10$^{46}$ erg/s. The source, however, is known to be magnified by a factor $\mu\sim$15-20 (Kormann et al. 1994; Assef et al. 2011). We can thus estimate the intrinsic luminosity to be L$_{2-10 keV}\sim$5$\times$10$^{45}$ erg s$^{-1}$ which translates to L$_{bol}\sim$4.5-9.4$\times$10$^{46}$ erg s$^{-1}$ once the bolometric correction ($k_{bol}$ ranges between $\sim$ 9-15 depending on the spectral index which, here, is found to range between $\Gamma \sim 1.65-1.95$ depending on the modeling of the data, Marchese et al. 2012) is also taken into account. This implies that the source is apparently emitting at $\sim$7-15\% of its Eddington luminosity ($L_{Edd}\sim$6.5$\times$10$^{47}$ erg s$^{-1}$). The measured 2-10 keV luminosity is rather typical of what is measured for lensed objects (Misawa et al. 2008) and it is also comparable to what is measured for a sample of radio quiet QSOs observed with XMM-Newton (L$_{2-10 keV}\sim$0.4-7$\times$10$^{46} erg s^{-1}$, Page et al. 2005) at redshift between z$\sim$2-4 and it is at the lowest end of the range of luminosities measured for radio loud objects (L$_{2-10 keV}\sim$0.5-70$\times$10$^{46} erg s^{-1}$, Page et al. 2005), suggesting that the source brightness is not experiencing, in X-rays, extreme boosting effects.  

The spectral analysis of the X-ray data clearly demonstrates that the spectrum of B1422+231 is complex and shows some highly significant features such as  a low-energy cut-off due to absorption, a deep dim of counts at E$\sim$7-7.5 keV (rest-frame) and a ``hump'' at E$\sim$ 20-30 keV plus a low energy cut-off.  We also have indications of the presence of an iron emission line at E$\sim$6.4 keV. All these evidences point towards the possible presence of a partial coverer (complex absorption scenario) and/or a reflection component (reflection scenario), i.e. towards the fact that we are observing a ``Seyfert-like'' X-ray spectrum. This scenario is further suggested by the detection of absorption in X-rays due to intervening matter as observed in nearby Seyferts. This component is apparently at odd with what predicted by the unified models of AGN (Antonucci 1993) when considering the presence of broad emission lines in the optical spectrum of B1422+231, and thus its "type I" nature is taken into account. It is worth considering, however, that also in the local Universe it is not unusual to observe type I Seyferts absorbed in X-rays (see for example the cases of NGC 4151 and of NGC 4051,  Warwick et al. 1993 and Ponti et al. 2006 respectively). In recent years, moreover, long term  monitoring of sources displaying characteristics of type I objects from optical-UV to X-rays for decades, may suddenly change their X-ray properties to that typical of type II objects, and maintain them for years (see for example the cases of H0557-385 and NGC 5548, Coffey et al. 2014; Kaastra et al. 2014). 

The complex X-ray spectrum of B1422+231, moreover, seems to be in contrast with the fact that the source has been defined to be radio-loud (Misawa et al. 2008) and, thus, a strong or even dominant jet component producing an almost featureless spectrum may be expected also in X-rays.  We thus investigated the radio-loudness of B1422+231 using the parameter defined by Kellermann et al. (1989), \hbox{$R$=$f_{\rm 5~GHz}/f_{\rm 4400~\mbox{\scriptsize\AA}}$}, where the optical wavelength and the radio frequency are in the source rest frame. Typical radio-loudness values are $>100$ for radio-loud quasars and $<10$ for radio-quiet quasars. The rest-frame 5~GHz flux density was computed from the 1.4 GHz  flux density value of 352~mJy reported by Tinti et al. (2005), extrapolated  to 5~GHz rest frame assuming a radio power-law slope of $\alpha=0.9$ as 
reported in Orienti et al. (2007), where \hbox{$S_{\nu}$ $\propto$ $\nu^{-\alpha}$}. The rest-frame 4400~\AA\ flux density was computed by adopting the 
composite quasar spectrum presented in Vanden Berk et al. (2001) to convert 
the broadband Sloan Digital Sky Survey (SDSS) $ugriz$ measurements to the flux 
density at the rest wavelength of 4400\AA\ (see $\S$2.2 of Vignali, Brandt 
and Schneider 2003 for further details). We obtain $R\approx90$, which indicates that B1422$+$231 is only moderately radio loud.

 On the other side, the rather-steep radio continuum measured by Orienti et al. (2007) with the VLA ($\alpha\sim$0.9) could indicate that the source is highly inclined in the plane of the sky. In this case the radio emission should be dominated by the lobes and not by the jet which is not pointing toward us, thus reconciling the radio- and X-ray properties of the target. Nonetheless VLBI imaging of B1422+231 does not allow the detection of any extended structures at more they $\sim$1-2 milliarcsec (Dallacasa, private communication). This means that the physical dimensions of the radio-source should be smaller than $\sim$10-20 kpc at the redshift of the source. We cannot probe to small enough scales to show that B1422+231 has intrinsic small radio-emitting regions such as those found in CoreG or FR0 sources (Baldi, Capetti \& Giovannini 2015) since the angular resolution of the VLBI is at the limit of the typical dimension of such sources. CoreG radio emitting regions have typical dimensions $\lesssim$10-20 kpc and we should be able to probe even smaller angular scales to exclude extended emission on the scales of the order $\sim$1-3 kpc that are typical of FR0 (Baldi, Capetti \& Giovannini 2015). Overall, the radio properties of B1422+231 prevent to firmly assess of the nature of the source. While the polarization in the radio-band and the radio flux variability, even if very small, may suggest the blazar nature of the source, the steady radio spectral shape and the low polarization at low frequencies point toward the fact that B1422+231 may be a genuinely young radio source (Orienti et al. 2007, Orienti \& Dallacasa 2008). To this evidence, we add the fact that, as most probably happens in the optical, also the X-ray band is dominated by the emission due to a disk component thus implying that the jet is not dominating at all frequencies. We finally note that also 4C74.26 the closest powerful radio-loud quasar displays a ``Seyfert-like'' X-ray spectrum (Grandi et al. 2006; Fukazawa et al. 2011).

\subsection{The two best-fit scenarios}

As mentioned above, the X-ray spectrum of B1422+231 is strongly characterized by the presence of features similar to those commonly observed in nearby Seyfert galaxies and possibly indicating that the primary emission is affected by complex absorption and/or reflection.

In the complex absorption scenario, we obtain that the primary engine is absorbed by a cold absorber with a column density of N$_H\sim$10$^{22}$ cm$^{-2}$ which completely covers the region were the X-ray photons originate plus a ionized ($\xi\sim$100 erg s$^{-1}$ cm and N$_H\sim$10$^{24}$ cm$^{-2}$) partial coverer that hides only $\sim$40\% of this region (C$_f\sim$0.4). 
Adopting this model, we obtain that the putative FeK$\alpha$ line is not detected at 99\% confidence level (EW$\lesssim$70 eV at 90\% confidence) but its strength is still consistent with what is expected for the observed absorbing column (Makishima 1986; Leahy \& Creighton, 1993; Matt, Perola \& Piro, 1991).  The partial covering of the central engine suggests that in the core of B1422+231 there are intervening clouds that are able to partially block the line of sight. This is similar to what is observed in nearby AGNs. In some Seyfert galaxies, clouds, probably the same forming the broad line regions, have been observed while crossing the line of sight and producing eclipses (Risaliti et al. 2007). Similarly, in other objects, the changes of absorbing column have been associated to outflowing clouds (Coffey et al. 2014; Kaastra et al. 2014; Arav et al. 2015). The changes in the absorption properties are thought to be responsible for at least part of the well known AGN variability seen in X-rays (see for example the case of NGC 1365, Risaliti et al. 2005, 2009;  see also Miller et al. 2008, 2010, Markowitz et al. 2014). In this scenario, assuming that the density of these clouds are typical for the broad line regions BLR (n$\sim$10$^{9}$ cm$^{-3}$) and taking the luminosity of the source obtained above ($\sim$5$\times$10$^{46}$ erg s$^{-1}$), it turns out that they are at d$\sim$7$\times$10$^{17}$ cm, corresponding to $\sim$5000r$_{g}$ or $\sim$0.1pc from  the source of X-ray photons, i.e. at a distance that is only $\sim$half of the expected average distance of the BLR for a QSO of the given L$_{bol}$ (in accordance to what found for the eclipsing cloud in NGC 1365 d$\sim$100r$_g$, Risaliti et al., 2007). An even looser limit is obtained following Tarter et al. (1969), under the assumption that d$_{max}\lesssim$L$_{ion}/\xi N_H$. If L$_{ion}$=L$_{2-10 keV}$, we obtain the d$_{max}\sim$4$\times$10$^{18}$ cm, i.e. one order of magnitude larger than what was obtained before. In both cases, the values obtained are in good agreement with results obtained by Tombesi et al (2012) in constraining the production region of UFOs detected in nearby AGN and by Markowitz et al. (2014) in constraining the distance of the clouds forming the torus. It is worth considering, however, that in this picture eclipsing episodes are responsible for the different observed X-ray flux states at which the sources are detected and that these episodes are generally related to cold clouds (Risaliti et al. 2007; Markowitz 2014; Coffey et al. 2014; Kaastra et al. 2015). Using archival data, we are unable to detect any variations of the absorbing column density. Moreover, the X-ray flux of B1422+231 observed during the 2014 observation is not significantly different from what previously measured, i.e. no long term variability has been detected. We also failed to detect significant short-term variability and no dominating flux dip is measured thus leaving us without indications of eclipsing episodes. It is worth considering, however, that the detection of strong variability, especially the short-time scale one, is here disfavored by both redshift of the source and by the mass of its SMBH (M$_{SMBH}\sim5\times$10$^{9}$M$_{\sun}$, Greene et al. 2010).  

Under the reflection scenario we obtained a value of $\chi^2$ almost identical to the one obtained with the complex absorption model but with a lower number of free parameters (see Table 2). The spectral index, once the spectral complexities have been taken into account with $Pexmon$, is found to be steep ($\Gamma\sim$1.8) and in agreement with what on average measured in the AGN in the nearby Universe (Perola et al. 2002; Cappi et al. 2006; Dadina 2008). This value is well within the range of photon index predicted for Seyfert galaxies (Haardt et al. 1994) and observed in QSO at intermediate and high redshift for objects with Eddington luminosity similar to that of B1422+231 (Brigthman et al. 2013). Most interestingly, the adoption of the $Pexmon$ model (Nandra et al. 2007) allows us to consistently fit the hump due to the reflection of the primary continuum onto the matter surrounding the central engine and the associated emission lines. In this framework, taking advantage of the source redshift and of the XMM-Newton sensitivity, we are able to measure both the high-energy cut-off (E$_c\sim$80 keV) and the normalization between the primary emission and the reflected component (r$\sim$1) that indicates that the reflecting matter covers $\sim$50\% of the sky as seen by the source of X-ray photons. These values are well within the range measured in nearby Seyferts (Perola et al. 2002; Dadina 2008). To our knowledge, this is the first AGN for which we are able to perform such measurements at z$\geq$3 (but see Lanzuisi et al. 2016 for the possible detection of such features in the X-ray spectrum of a z$\sim$2 QSO). This allows us to probe the geometry of the accreting matter around a SMBH at ages when the QSOs activity was at its peak, thus possibly opening new opportunities to test the SMBH-galaxy feedback and formation models (King \& Pounds 2015 and references therein). The detection of the high-energy cut-off, moreover, allows us to infer the temperature of the putative Comptonizing corona (kT$\sim \frac{E_{cut-off}}{2}$) and its optical depth that, following eq.1 in Petrucci et al. (2001) is $\tau\sim$0.3 for B1422+231. It is worth noting, however, that the kT values derived from $Pexmon$, that is derived from $Pexrav$ (Magdziarz \& Zdziarski 1995) must be considered to be, in general, under-estimated and $\tau$ over-estimated. This is because of the spherical and isotropic geometry assumed in this model. A more realistic slab and anisotropic geometry of Comptonization model should drive towards different values of E$_{c}$ and $\tau$ and the value presented above must be taken with cautions.

\section{Summary and conclusions}

We can summarize our findings as follow: 1) the coupling of the magnification factor due to gravitational lensing and the sensitivity of XMM-Newton allowed us to obtain a very high-quality X-ray spectrum of a QSO at z=3.62. This technique is thus proved to be useful to directly probe the physical properties of the matter surrounding the SMBH at high redshift, i.e. at the peak of both QSO and starburst activities; 2) the source X-ray spectrum is certainly complex displaying strong and highly significant signatures of both emission and absorption components; 3) the spectrum is equally well fitted by either a complex absorption model and a reflection  model.  

Even if we have no firm indication of which of the two best-fit models proposed here is the right one, we favor the reflection dominated interpretation because a) we do not detect any hint of variability both on short and long time scale so that the presence of passing and partially absorbing clouds seems to be disfavored and b) the reflection model is better on a pure statistical basis allowing to obtain a good modeling of the data with fewer parameters.  In this scenario, we stress that we have been able to measure the strength of the reflection component (r$\sim$1) and the temperature of the Comptonizing corona inferring that the geometry and the physical state of the matter around the SMBH hosted in the z=3.62 B1422+231 QSO are similar to what measured in nearby Seyfert galaxies.

\begin{acknowledgements} We thank the anonymous referee for her/his helpful comments that helped to improve the manuscript. This work is based on observations obtained with XMM-Newton, an ESA science mission with instruments and contributions directly funded by ESA Member States and the USA (NASA). MD thanks Paola Grandi, Daniele Dallacasa for useful discussions. MD and MC aknowledge financial support from the Italian Space Agency under grant ASI-INAF I/037/12/0 and extended PRIN INAF 2012. GL acknowledges financial support from the CIG grant ``eEASY'' n. 321913, from ASI-INAF 2014-045-R.0 and ASI/INAF I/037/12/0–011/13 grants. GP thanks the DLR: the Bundesministerium f\"ur Wirtschaft und Technologie/Deutsche Zentrum f\"ur luft-un Raumfahrt (BMWI/DLR, FKZ 50 OR 1408). 

\end{acknowledgements}

%-------------------------------------------------------------------

{}


\begin{thebibliography}{}

\bibitem[Aird et al.(2015)]{2015MNRAS.451.1892A} Aird, J., Coil, A.~L., Georgakakis, A., et al.\ 2015, \mnras, 451, 1892 

\bibitem[Antonucci(1993)]{1993ARA&A..31..473A} Antonucci, R.\ 1993, \araa, 31, 473 

\bibitem[Arav et 
al.(2015)]{2015A&A...577A..37A} Arav, N., Chamberlain, C., Kriss, G.~A., et al.\ 2015, \aap, 577, A37 

\bibitem[Assef et al.(2011)]{2011ApJ...742...93A} Assef, R.~J., Denney, K.~D., Kochanek, C.~S., et al.\ 2011, \apj, 742, 93 

\bibitem[Baldi et 
al.(2015)]{2015A&A...576A..38B} Baldi, R.~D., Capetti, A., \& Giovannini, G.\ 2015, \aap, 576, A38 


\bibitem[Bassani et al.(1999)]{1999ApJS..121..473B} Bassani, L., Dadina, 
M., Maiolino, R., et al.\ 1999, \apjs, 121, 473 

\bibitem[Boyle et al.(1988)]{1988MNRAS.235..935B} Boyle, B.~J., Shanks, T., 
\& Peterson, B.~A.\ 1988, \mnras, 235, 935 

\bibitem[Brandt 
\& Hasinger(2005)]{2005ARA&A..43..827B} Brandt, W.~N., \& Hasinger, G.\ 2005, \araa, 43, 827 

\bibitem[Brightman et al.(2013)]{2013MNRAS.433.2485B} Brightman, M., 
Silverman, J.~D., Mainieri, V., et al.\ 2013, \mnras, 433, 2485

\bibitem[Cappi et al.(2006)]{2006A&A...446..459C} Cappi, M., Panessa, F., Bassani, L., et al.\ 2006, \aap, 446, 459 

\bibitem[Cartas et al. (2016)]{} Chartas et al. , 2016, Astron. Nach. AN 337, No. 4/5, 356-361

\bibitem[Chartas et al.(2002)]{2002ApJ...579..169C} Chartas, G., Brandt, 
W.~N., Gallagher, S.~C., \& Garmire, G.~P.\ 2002, \apj, 579, 169 

\bibitem[Chartas et al.(2003)]{2003ApJ...595...85C} Chartas, G., Brandt, 
W.~N., \& Gallagher, S.~C.\ 2003, \apj, 595, 85

\bibitem[Chartas et al.(2007)]{2007ApJ...661..678C} Chartas, G., Eracleous, 
M., Dai, X., Agol, E., \& Gallagher, S.\ 2007, \apj, 661, 678 

\bibitem[Chartas et al.(2014)]{2014ApJ...783...57C} Chartas, G., Hamann, 
F., Eracleous, M., et al.\ 2014, \apj, 783, 57 

\bibitem[Chiba(2002)]{2002ApJ...565...17C} Chiba, M.\ 2002, \apj, 565, 17 

\bibitem[Coffey et al.(2014)]{2014MNRAS.443.1788C} Coffey, D., Longinotti, 
A.~L., Rodr{\'{\i}}guez-Ardila, A., et al.\ 2014, \mnras, 443, 1788 


\bibitem[Dadina et al.(2005)]{2005A&A...442..461D} Dadina, M., Cappi, M., Malaguti, G., Ponti, G., \& de Rosa, A.\ 2005, \aap, 442, 461 

\bibitem[Dadina(2007)]{2007A&A...461.1209D} Dadina, M.\ 2007, \aap, 461, 1209 

\bibitem[Dadina(2008)]{2008A&A...485..417D} Dadina, M.\ 2008, \aap, 485, 417 

\bibitem[de La Calle P{\'e}rez et 
al.(2010)]{2010A&A...524A..50D} de La Calle P{\'e}rez, I., Longinotti, A.~L., Guainazzi, M., et al.\ 2010, \aap, 524, A50 

\bibitem[Done et al.(2007)]{2007MNRAS.374L..15D} Done, C., Sobolewska, 
M.~A., Gierli{\'n}ski, M., \& Schurch, N.~J.\ 2007, \mnras, 374, L15 

\bibitem[Fabian et al.(2000)]{2000PASP..112.1145F} Fabian, A.~C., Iwasawa, 
K., Reynolds, C.~S., \& Young, A.~J.\ 2000, \pasp, 112, 1145 


\bibitem[Fabian et al.(2015)]{2015MNRAS.451.4375F} Fabian, A.~C., Lohfink, 
A., Kara, E., et al.\ 2015, \mnras, 451, 4375 

\bibitem[Fanali et al.(2013)]{2013MNRAS.433..648F} Fanali, R., Caccianiga, 
A., Severgnini, P., et al.\ 2013, \mnras, 433, 648 

\bibitem[Ferrarese \& Merritt(2000)]{2000ApJ...539L...9F} Ferrarese, L., \& Merritt, D.\ 2000, \apjl, 539, L9 

\bibitem[Fukazawa et al.(2011)]{2011ApJ...727...19F} Fukazawa, Y., Hiragi, K., Mizuno, M., et al.\ 2011, \apj, 727, 19 

\bibitem[Garc{\'{\i}}a et al.(2014)]{2014ApJ...782...76G} Garc{\'{\i}}a, J., Dauser, T., Lohfink, A., et al.\ 2014, \apj, 782, 76 

\bibitem[Gofford et al.(2013)]{2013MNRAS.430...60G} Gofford, J., Reeves, 
J.~N., Tombesi, F., et al.\ 2013, \mnras, 430, 60

\bibitem[Grandi et al.(2006)]{2006ApJ...642..113G} Grandi, P., Malaguti, G., \& Fiocchi, M.\ 2006, \apj, 642, 113 

 \bibitem[Grandi \& Palumbo(2004)]{2004Sci...306..998G} Grandi, P., \& Palumbo, G.~G.~C.\ 2004, Science, 306, 998 

\bibitem[Grandi 
\& Palumbo(2007)]{2007ApJ...659..235G} Grandi, P., \& Palumbo, G.~G.~C.\ 2007, \apj, 659, 235 


\bibitem[Greene et al.(2010)]{2010ApJ...709..937G} Greene, J.~E., Peng, C.~Y., \& Ludwig, R.~R.\ 2010, \apj, 709, 937

\bibitem[Guilbert 
\& Rees(1988)]{1988MNRAS.233..475G} Guilbert, P.~W., \& Rees, M.~J.\ 1988, \mnras, 233, 475 


\bibitem[Haardt 
\& Maraschi(1993)]{1993ApJ...413..507H} Haardt, F., \& Maraschi, L.\ 1993, \apj, 413, 507 


\bibitem[Haardt et al.(1994)]{1994ApJ...432L..95H} Haardt, F., Maraschi, 
L., \& Ghisellini, G.\ 1994, \apjl, 432, L95 

\bibitem[Haardt et al.(1997)]{1997ApJ...476..620H} Haardt, F., Maraschi, 
L., \& Ghisellini, G.\ 1997, \apj, 476, 620 



\bibitem[Kaastra et al.(2014)]{2014Sci...345...64K} Kaastra, J.~S., Kriss, 
G.~A., Cappi, M., et al.\ 2014, Science, 345, 64 

\bibitem[Kalberla et 
al.(2005)]{2005A&A...440..775K} Kalberla, P.~M.~W., Burton, W.~B., Hartmann, D., et al.\ 2005, \aap, 440, 775 



\bibitem[Kellermann et al.(1989)]{1989AJ.....98.1195K} Kellermann, K.~I., 
Sramek, R., Schmidt, M., Shaffer, D.~B., \& Green, R.\ 1989, \aj, 98, 1195 

\bibitem[King 
\& Pounds(2015)]{2015ARA&A..53..115K} King, A., \& Pounds, K.\ 2015, \araa, 53, 115 


\bibitem[Kormann et al.(1994)]{1994A&A...286..357K} Kormann, R., Schneider, P., \& Bartelmann, M.\ 1994, \aap, 286, 357 

\bibitem[Kormendy \& Richstone(1995)]{1995ARA&A..33..581K} Kormendy, J., \& Richstone, D.\ 1995, \araa, 33, 581 


\bibitem[Lanzuisi et al.(2012)]{2012A&A...544A...2L} Lanzuisi, G., Giustini, M., Cappi, M., et al.\ 2012, \aap, 544, A2 

\bibitem[Lanzuisi et al.(2016)]{2016arXiv160402462L} Lanzuisi, G., Perna, M., Comastri, A., et al.\ 2016, arXiv:1604.02462 

\bibitem[Leahy 
\& Creighton(1993)]{1993MNRAS.263..314L} Leahy, D.~A., \& Creighton, J.\ 1993, \mnras, 263, 314 

\bibitem[Lightman 
\& White(1988)]{1988ApJ...335...57L} Lightman, A.~P., \& White, T.~R.\ 1988, \apj, 335, 57 


\bibitem[Lu 
\& Yu(1999)]{1999ApJ...526L...5L} Lu, Y., \& Yu, Q.\ 1999, \apjl, 526, L5 

\bibitem[Madau et al.(1996)]{1996MNRAS.283.1388M} Madau, P., Ferguson, 
H.~C., Dickinson, M.~E., et al.\ 1996, \mnras, 283, 1388 

\bibitem[Magdziarz 
\& Zdziarski(1995)]{1995MNRAS.273..837M} Magdziarz, P., \& Zdziarski, A.~A.\ 1995, \mnras, 273, 837 

\bibitem[Magorrian et al.(1998)]{1998AJ....115.2285M} Magorrian, J., 
Tremaine, S., Richstone, D., et al.\ 1998, \aj, 115, 2285 


\bibitem[Makishima(1986)]{1986LNP...266..249M} Makishima, K.\ 1986, The 
Physics of Accretion onto Compact Objects, 266, 249 

\bibitem[Malizia et al.(2014)]{2014ApJ...782L..25M} Malizia, A., Molina, 
M., Bassani, L., et al.\ 2014, \apjl, 782, L25 

\bibitem[Marchese et 
al.(2012)]{2012A&A...539A..48M} Marchese, E., Della Ceca, R., Caccianiga, A., et al.\ 2012, \aap, 539, A48 

\bibitem[Marconi et al.(2004)]{2004MNRAS.351..169M} Marconi, A., Risaliti, G., Gilli, R., et al.\ 2004, \mnras, 351, 169 

\bibitem[Markowitz et al.(2014)]{2014MNRAS.439.1403M} Markowitz, A.~G., 
Krumpe, M., \& Nikutta, R.\ 2014, \mnras, 439, 1403 


\bibitem[Matt et 
al.(1991)]{1991A&A...247...25M} Matt, G., Perola, G.~C., \& Piro, L.\ 1991, \aap, 247, 25 

\bibitem[Merritt \& Ferrarese(2001)]{2001MNRAS.320L..30M} Merritt, D., \& Ferrarese, L.\ 2001, \mnras, 320, L30 

\bibitem[Miller et 
al.(2008)]{2008A&A...483..437M} Miller, L., Turner, T.~J., \& Reeves, J.~N.\ 2008, \aap, 483, 437 

\bibitem[Miller et al.(2010)]{2010MNRAS.408.1928M} Miller, L., Turner, 
T.~J., Reeves, J.~N., \& Braito, V.\ 2010, \mnras, 408, 1928

\bibitem[Misawa et al.(2007)]{2007ApJS..171....1M} Misawa, T., Charlton, 
J.~C., Eracleous, M., et al.\ 2007, \apjs, 171, 1 

\bibitem[Misawa et al.(2008)]{2008ApJ...677..863M} Misawa, T., Eracleous, 
M., Chartas, G., \& Charlton, J.~C.\ 2008, \apj, 677, 863 

\bibitem[Murphy 
\& Yaqoob(2009)]{2009MNRAS.397.1549M} Murphy, K.~D., \& Yaqoob, T.\ 2009, \mnras, 397, 1549 

\bibitem[Nandra et al.(2013)]{2013arXiv1306.2307N} Nandra, K., Barret, D., 
Barcons, X., et al.\ 2013, arXiv:1306.2307 

\bibitem[Nandra et al.(2007)]{2007MNRAS.382..194N} Nandra, K., O'Neill, 
P.~M., George, I.~M., \& Reeves, J.~N.\ 2007, \mnras, 382, 194 


\bibitem[Nandra \& Pounds(1994)]{1994MNRAS.268..405N} Nandra, K., \& Pounds, K.~A.\ 1994, \mnras, 268, 405



\bibitem[Nardini et al.(2015)]{2015Sci...347..860N} Nardini, E., Reeves, 
J.~N., Gofford, J., et al.\ 2015, Science, 347, 860 

\bibitem[Orienti 
\& Dallacasa(2008)]{2008A&A...479..409O} Orienti, M., \& Dallacasa, D.\ 2008, \aap, 479, 409 


\bibitem[Orienti et 
al.(2007)]{2007A&A...475..813O} Orienti, M., Dallacasa, D., \& Stanghellini, C.\ 2007, \aap, 475, 813 


\bibitem[Paczynski(1986)]{1986ApJ...301..503P} Paczynski, B.\ 1986, \apj, 
301, 503 

\bibitem[Page et al.(2005)]{2005MNRAS.364..195P} Page, K.~L., Reeves, 
J.~N., O'Brien, P.~T., \& Turner, M.~J.~L.\ 2005, \mnras, 364, 195 

\bibitem[Patnaik et al.(1992)]{1992MNRAS.259P...1P} Patnaik, A.~R., Browne, 
I.~W.~A., Walsh, D., Chaffee, F.~H., \& Foltz, C.~B.\ 1992, \mnras, 259, 1P 

\bibitem[Perola et 
al.(2002)]{2002A&A...389..802P} Perola, G.~C., Matt, G., Cappi, M., et al.\ 2002, \aap, 389, 802 

\bibitem[Petrucci et al.(2001)]{2001ApJ...556..716P} Petrucci, P.~O., 
Haardt, F., Maraschi, L., et al.\ 2001, \apj, 556, 716 

\bibitem[Piconcelli et al.(2005)]{2005A&A...432...15P} Piconcelli, E., Jimenez-Bail{\'o}n, E., Guainazzi, M., et al.\ 2005, \aap, 432, 15 

\bibitem[Protassov et al.(2002)]{2002ApJ...571..545P} Protassov, R., van Dyk, D.~A., Connors, A., Kashyap, V.~L., \& Siemiginowska, A.\ 2002, \apj, 571, 545 

\bibitem[Ponti et al.(2006)]{2006AN....327.1055P} Ponti, G., Miniutti, G., Fabian, A.~C., Cappi, M., \& Palumbo, G.~G.~C.\ 2006, Astronomische Nachrichten, 327, 1055 

\bibitem[Pounds et al.(1995)]{1995MNRAS.277L...5P} Pounds, K.~A., Done, C., 
\& Osborne, J.~P.\ 1995, \mnras, 277, L5 

\bibitem[Pounds et al.(2003)]{2003MNRAS.345..705P} Pounds, K.~A., Reeves, 
J.~N., King, A.~R., et al.\ 2003, \mnras, 345, 705 

\bibitem[Raychaudhury et al.(2003)]{2003AJ....126...29R} Raychaudhury, S., Saha, P., \& Williams, L.~L.~R.\ 2003, \aj, 126, 29 

\bibitem[Reeves et al.(2008)]{2008MNRAS.385L.108R} Reeves, J., Done, C., 
Pounds, K., et al.\ 2008, \mnras, 385, L108 

\bibitem[Reis et al.(2014)]{2014Natur.507..207R} Reis, R.~C., Reynolds, 
M.~T., Miller, J.~M., \& Walton, D.~J.\ 2014, \nat, 507, 207 

\bibitem[Reynolds et al.(2014)]{2014ApJ...792L..19R} Reynolds, M.~T., 
Walton, D.~J., Miller, J.~M., \& Reis, R.~C.\ 2014, \apjl, 792, L19 

\bibitem[Risaliti et al.(2005)]{2005ApJ...623L..93R} Risaliti, G., Elvis, 
M., Fabbiano, G., Baldi, A., \& Zezas, A.\ 2005, \apjl, 623, L93 

\bibitem[Risaliti et al.(2007)]{2007ApJ...659L.111R} Risaliti, G., Elvis, 
M., Fabbiano, G., et al.\ 2007, \apjl, 659, L111 


\bibitem[Risaliti et al.(2009)]{2009ApJ...700L...6R} Risaliti, G., Young, 
M., \& Elvis, M.\ 2009, \apjl, 700, L6 


\bibitem[Saez \& Chartas(2011)]{2011ApJ...737...91S} Saez, C., \& Chartas, G.\ 2011, \apj, 737, 91 

\bibitem[Saez et al. (2008)]{} Saez, C., Chartas, G., Brandt, W.~N., et al.\ 2008, \aj, 135, 1505

\bibitem[Shemmer et al.(2006)]{2006ApJ...646L..29S} Shemmer, O., Brandt, 
W.~N., Netzer, H., Maiolino, R., \& Kaspi, S.\ 2006, \apjl, 646, L29 

\bibitem[Shemmer et al.(2008)]{2008ApJ...682...81S} Shemmer, O., Brandt, 
W.~N., Netzer, H., Maiolino, R., \& Kaspi, S.\ 2008, \apj, 682, 81 

\bibitem[Smith \& Done(1996)]{1996MNRAS.280..355S} Smith, D.~A., \& Done, C.\ 1996, \mnras, 280, 355 

\bibitem[Tanaka et al.(1995)]{1995Natur.375..659T} Tanaka, Y., Nandra, K., 
Fabian, A.~C., et al.\ 1995, \nat, 375, 659 

\bibitem[Tarter 
\& Salpeter(1969)]{1969ApJ...156..953T} Tarter, C.~B., \& Salpeter, E.~E.\ 1969, \apj, 156, 953 


\bibitem[Tinti et 
al.(2005)]{2005A&A...432...31T} Tinti, S., Dallacasa, D., de Zotti, G., Celotti, A., \& Stanghellini, C.\ 2005, \aap, 432, 31 


\bibitem[Tombesi et al.(2010)]{2010A&A...521A..57T} Tombesi, F., Cappi, M., Reeves, J.~N., et al.\ 2010a, \aap, 521

\bibitem[Tombesi et al.(2010)]{2010ApJ...719..700T} Tombesi, F., Sambruna, 
R.~M., Reeves, J.~N., et al.\ 2010b, \apj, 719, 700 

\bibitem[Vanden Berk et al.(2001)]{2001AJ....122..549V} Vanden Berk, D.~E., 
Richards, G.~T., Bauer, A., et al.\ 2001, \aj, 122, 549

\bibitem[Vignali et al.(2006)]{2006MNRAS.373..321V} Vignali, C., Alexander, 
D.~M., \& Comastri, A.\ 2006, \mnras, 373, 321 

\bibitem[Vignali et al.(2010)]{2010MNRAS.404...48V} Vignali, C., Alexander, 
D.~M., Gilli, R., \& Pozzi, F.\ 2010, \mnras, 404, 48 

\bibitem[Vignali et al.(2003)]{2003AJ....125..433V} Vignali, C., Brandt, 
W.~N., \& Schneider, D.~P.\ 2003, \aj, 125, 433 

\bibitem[Vignali et al.(2005)]{2005AJ....129.2519V} Vignali, C., Brandt, 
W.~N., Schneider, D.~P., \& Kaspi, S.\ 2005, \aj, 129, 2519 

\bibitem[Vignali et 
al.(2015)]{2015A&A...583A.141V} Vignali, C., Iwasawa, K., Comastri, A., et al.\ 2015, \aap, 583, A141 

\bibitem[Vignali et 
al.(2014)]{2014A&A...571A..34V} Vignali, C., Mignoli, M., Gilli, R., et al.\ 2014, \aap, 571, A34 

\bibitem[Walton et al.(2015)]{2015ApJ...805..161W} Walton, D.~J., Reynolds, 
M.~T., Miller, J.~M., et al.\ 2015, \apj, 805, 161 

\bibitem[Wambsganss(1990)]{1990PhDT.......180W} Wambsganss, J.\ 1990, 
Ph.D.~Thesis, 

\bibitem[Warwick et al.(1993)]{1993MNRAS.265..412W} Warwick, R.~S., Sembay, S., Yaqoob, T., et al.\ 1993, \mnras, 265, 412 

\end{thebibliography}
\end{document}